# Vitaly Ginzburg and High Temperature Superconductivity: Personal Reminiscences.

> 'АЩЕ ЇЕСМЬ НЕ МУДРЪ НО ВЪ ПР МУДРИХЪ РИЗУ ОБЛАЧИХСА А СМ ЫСЛЕННИХЪ САПОГИ НОСИЛЪ ЇЕСМЬ
>
> *As coarse as my own wits are, I have been privileged to wear the boots of the Sagacious and be clad in the robes of the Virtuous*
>
> *The Supplication of Daniel the Prisoner, Russia, XIII Century*

For this volume, with the permission of the Editors, rather than discussing Room Temperature Superconductivity *per se,* I have elected to share with the readers my reminiscences from the period of 1976-1983, when I was a M. Sc. and then a Ph.D. student in Vitaly L. Ginzburg's High Temperature Superconductivity group at the P.N. Lebedev Institute in Moscow.

I think I need to start first with some background information on the narrator, the time and the place, which should help the reader to properly place my story.

I graduated from high school in 1971, and, at only 16 years of age, was presented with the tough choice of a future career. In the Soviet Universities, applicants were to declare their major *before* their freshman year, not as sophomores, as in the U.S.[1] In my case, I was vacillating between physics and linguistics, and the choice was, ironically, determined by the Soviet system itself: physics was taught at many colleges, not only at top league schools like Moscow State or Moscow Institute for Science and Technology (MIPT), but linguistics was offered essentially only at the Department of Philology of the MSU, and hardly any student of Jewish background would be admitted there in the early 70's[2]. So, I was sort of predestined to try my hand at physics.

The Soviet admission system was, theoretically, in many aspects superior to the American one. Instead of relying on school-dependent GPAs and ill-defined aptitude tests, all colleges administered entrance tests in the leading subjects (for physics and technology this would usually be physics and math). In principle, this system allowed for a thorough selection of gifted students. The famed Institute for Science and Technology was administering 4 tests, a written and an oral examination in both math and physics. As opposed to the Physics Department of MSU, the administration of MIPT was not anti-Semitic, so their actual policy varied depending on how much pressure was exercised in a given year from the Party authorities. My year, 1971, was particularly bad. A strict 2% limit on Jewish admission was applied. Admission to the "faculty" (division) I was applying to, the Faculty of Chemical and Molecular Physics, was 100 persons. The top score of 20 points was not reached by any applicant, 19 points by two, one of them a Jew, Michael Feigel'man (who eventually got admitted and is now the Deputy Director of the Landau Institute and the Chair of the Theoretical Physics Department at MIPT), 18 by approximately 15 kids, including myself and 3 other Jews. Admitting all four of us would have been a violation of the unofficial limit of 2%, selecting one of four was too much of a hassle, so all four were rejected.

Many second tier colleges were not subject to such close monitoring by the authorities, and some of them chose to profit from the Party policy. A number of unofficial recruiters from various technical colleges were hanging around the MIPT campus, hinting to rejected applicants that their institutions may be willing to waive entrance tests if the score from the MIPT tests were high enough. One of such recruiters from the Moscow Institute of Steel and Alloys picked up me and a number of other Jews with 17 and 18 point. This is how I got to this college that was mostly engineering, but also graduated a score of metal physicists every year. This is how my destiny as a solid state physicist was fixed.

In the four years that followed I hardly had any choices: there were no electives in the Soviet universities, and every single course was a requirement. The next time I was presented with freedom of choice was when I was up for my Master thesis work in 1976, and by this time I knew that I wanted to become a theorist. The two main centers for theoretical physics in Moscow were the Landau Institute, in the small town of Chernogolovka about an hour away from the city, and the Theoretical Department of the Lebedev Institute, incidentally, within walking distance from my home. The latter was led by Vitaly Lazarevich Ginzburg, known at the Institute under the nickname "V.L."[3], who was at that time, a decade after Landau's death, generally considered to be the patriarch of Soviet theoretical physics. So, I asked for permission from my *alma mater* to work on my thesis at Lebedev, and permission was granted both to me and to a friend of mine, another aspiring theoretical physicist who was also rejected by a top-rank university (in his case MSU) and also picked up by recruiters from the Steel Institute. His name was Alex Gurevich and he is now one of the leading theorists in the U.S. in applied superconductivity. Alex and I showed up in Ginzburg's department one morning and declared our desire to work on our theses under the guidance of prominent Lebedev theorists. Curiously, we were not rejected, and in fact were both assigned problems related to high temperature superconductivity.

What we did not know was that a few years earlier V.L. had successfully lobbied the Academy of Science of the USSR for funding the dream of his life - a quest for high temperature superconductivity. I should add that V.L. is of the true enthusiasts' ilk; his enthusiasm was contagious, infectious. It was not that easy to warm up the Academy bureaucrats, but V.L.'s inner energy was overwhelming. He got the money, and got enough to hire several outstanding people. To name a few, among the new hires were Lev Bulaevsky, now at Los Alamos, Daniel Khomskii, now at the University of Cologne, and Andrey Linde, now at Stanford, all of them renowned leaders in their respective fields of research.

The principal questions formulated by Ginzburg for the newly created Superconductivity Section were (i) are there any principal limitations on the superconducting transition temperature that would prevent HTSC or RTSC at all or for the phonon mechanism, (ii) what are the most promising routes to enhance $T_c$ in conventional materials and (iii) what are alternative mechanisms that would lead to a radical improvement of superconducting properties.

The first question seems futile now, 20 years after the discovery of the HTSC, but at some time it was considered by many a requirement for the complete theory of superconductivity to predict a "sensible" (that is, around 30 K) limit on superconducting temperature[4]. Moreover, this prediction was actually made in 1972 by Marvin Cohen and Phillip Anderson[5] (they did not know that essentially the same argument had been examined two years earlier by Kirzhnits, Maksimov and Khomskii[6] because an English translation appeared only three years later). The argument, in a nutshell (using modern ideas about strong coupling and the McMillan equation, but keeping the original physics), goes like this: Crystal stability requires the static dielectric function, $\varepsilon(0,\mathbf{q})$ to be positive, otherwise a spontaneous charge density wave will be generated at the corresponding $\mathbf{q}$. This, roughly speaking, leads to a condition $\lambda<\mu$, where $\lambda$ is the constant characterizing coupling with bosons responsible for the pairing attraction (in conventional systems, phonons), and $\mu$ is the corresponding constant for the Coulomb repulsion. This means that superconductivity is only possible because, as was first realized by Tolmachev[7] and independently by Morel and Anderson[8], the Coulomb repulsion is logarithmically renormalized due to the large difference in energy scales. The standard simplified expression for $T_c$ is then
$$T_c \approx \theta \exp\{-1/[\lambda/(1+\lambda)-\mu^*]\},$$
where $\theta$ is the characteristic energy of the intermediate bosons, and $\mu^*=\mu/[1+\mu \ln(E/\theta)]$, E being the energy scale of the Coulomb repulsion. Assuming $\lambda=\mu$ one can optimize this expression with respect to $\theta$ and the resulting temperature, $E \exp(-4-3/\lambda)$ is only a few tens of Kelvins for any reasonable values of E and $\lambda$[9].

This argument is, of course, flawed, and not only because of the (seemingly unimportant at the time) assumptions of uniformity and isotropy of both electronic structure and superconducting gap (both are severely violated, for instance, in $MgB_2$), but because of a principal mistake, pointed out by Kirzhnits and his co-workers[10]: the correct condition for system stability is not $\varepsilon(0,\mathbf{q})>0$, but $\varepsilon^{-1}(0,\mathbf{q})<1$, which allows for *negative* values of $\varepsilon$ (but not values between 0 and 1). This was shown rigorously by David Kirzhnits[11] in 1976[12], but he initially assumed that a negative $\varepsilon$ was just an abstract possibility. The beauty of Ref. 10 was that it showed that $\varepsilon(0,\mathbf{q})<0$ at some $\mathbf{q}$'s in many regular systems.

With the idea of a principal limit on $T_c$ out of the way, V.L. was encouraging his newly formed group to exploit different avenues potentially leading to HTSC. While the conventional route was by no means abandoned, V.L. himself strongly favored electronic mechanisms (he used to call it "excitonic superconductivity", meaning, however, arbitrary pairing interaction of a non-phonon origin).

It is curious how two great minds in the theory of superconductivity have both missed an opportunity that with the benefit of hindsight seems nearly obvious. Both V.L. and P.W. Anderson rejected the possibility of building HTSC based on repulsive interactions. In fact, Anderson writes explicitly in Ref. 4: "In most of the more complicated mechanisms, electrons seem to be paired in anisotropic or otherwise unusual states, which are broken up by impurity scattering". At that time V.L. would probably disagree with the first part (his bet was always with an s-wave pairing mediated by attractive interactions of electronic origin), but would agree with the second: that anisotropic and, in particular, non-s-wave states would be destroyed by impurities. It has turned out that Anderson was right on the first count, both cuprates and $MgB_2$ have highly anisotropic order parameters, but was wrong on the second. Implicitly, he had in mind moderately high critical temperatures, but in reality, the higher $T_c$, the larger the gap, $\Delta$, and the shorter the coherence length, $\xi$, (inversely proportional to the gap value), so that for *really high* $T_c$ the condition for purity, $l_{m.f.p.}>\xi$ appears to be very mild. Anderson was right also on another count, when he said later in the same article that there is some "other mechanism" that will probably occur, because "the requirement … is that the interaction be attractive not everywhere but simply in some, not necessarily very large, region of space in time"[13] (admittedly, he later predicts that "the transition temperature would be exponentially low"…).

Anyway, the idea of d-wave pairing was not on the table in Ginzburg's group. However, many of the areas in which efforts were concentrated have proven to be very fruitful later, such as superconductivity in low-dimensional structures (layered and quasi-1D organic), mostly pursued by Lev Boulaevskii, the interplay between excitonic insulators and superconductivity, covered by Yury Kopaev (some of his results were rediscovered 20 years later in connection to hexaborides[14]), "sandwich" superconductivity (Zharkov and Uspenski), also known as the "ginzburger", and nonequilibrium superconductivity, also studied by Kopaev. In fact, many of the results obtained at that time were not appreciated by later researchers. For instance, the infamous field-effect superconductivity, "discovered" in 2000 by Hendrik Schön, essentially builds upon the above study of Kopaev. Some other unfairly forgotten, but useful and important results include Maksimov's formulation of the Eliashberg theory in real space (from which it follows, in particular, that even for a spatially inhomogeneous system the total electron-phonon coupling constant is strictly independent of ionic masses, a rigorous, but little known result) or Uspenski's study of the lattice stability restrictions on s-wave superconductivity of electronic origin[15] that that I will come back to once more later in the article.

Much of this effort was summarized in a book, published in Russian in 1977 under the title *The problem of high-temperature superconductivity*.[16] The manuscript was mostly finished by 1976, when I first showed up at the Theoretical Department, expressing my wish to work on my Master's thesis in Ginzburg's group. I was offered the possibility to work under the supervision of Eugene Maksimov, who suggested as a main topic superconductivity mediated by acoustic plasmons. Such a possibility was proposed in 1966-68 by several authors, including Herbert Frölich in Liverpool (who was, as usual, not aware that he was replicating an older paper by Geilikman) and Eduard Pashitski[17] in Kiev), but probably mostly elaborated by Geilikman and his collaborators in Moscow[18]. Maksimov's proposal was to marry this idea with the "ginzburger" concept in an imaginary metal consisting of two sets of relatively flat bands separated by a small energy gap, and a light band crossing both (with only light electrons at $E_F$). It turned out, however, that not only this was not a viable model, but in fact the previously discussed acoustic plasmon mechanism was even less viable. The point, missed in the previous works, as well as in subsequent ones (proposals of acoustic plasmon superconductivity keep popping up quite regularly for every new superconductor discovered: cuprates, fullerites, $MgB_2$, $CaC_6$… though surprisingly, nobody has yet summoned acoustic plasmons for the cobaltate, but that may be coming), was that while for the total dielectric function the stability condition $\varepsilon(0,\mathbf{q})>0$ does not hold, for its *electronic* part it does hold. This is very clear from the two relevant diagrams:

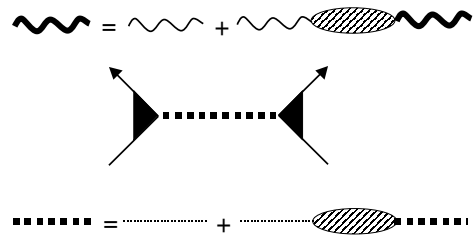

The top diagram describes renormalization of the phonon frequency due to screening by electrons. The full electronic susceptibility, the hatched bubble, cannot be negative since this would render the phonons unstable. Therefore, the effective electron-electron interaction due to plasmons, the second diagram, can only become attractive because of local field effects, represented by the vertexes in the second diagram. The thick broken line there, representing the plasmon propagator, cannot be negative, because it is defined (see the last diagram) by the same equation as the phonon renormalization. This does not render plasmon superconductivity impossible, but severely restricts the acceptable range of parameters. A detailed discussion of this can be found in Ref. 15.

Let me at this point digress from physics and describe the atmosphere at the Theoretical department as it opened up to me in 1976. Ginzburg's credo[19] (that he shared with his older colleague, Landau) was that a theorist cannot be limited by one field or even subfield[20]. He himself was active in condensed matter, in quantum field theory, and in astrophysics. The rule, unwritten, but strictly required by V.L., was that every person affiliated with the Department, be it a member of the Academy or graduate student, was to attend two seminars a week[21], one of which must be one of the two "big" seminars. One of these was moderated by Efim Fradkin (known in the solid state community for his invention of the temperature Green functions), and catered mostly to the high energy/quantum field crowd. The other was the gargantuan Moscow Seminar on Theoretical Physics, moderated by V.L. himself and attracting 200-300 participants every week[22]. The seminar was famous among Moscow physicists. I recall that after the 1000$^{th}$ seminar that was filled with jocular talks, a tape of a made-up "street recording" was played in which a tourist asked for the way to a department store, and a housewife in the street explained that it could be found across the street from Ginzburg's seminar.

The other "required" seminar would have been a specialized seminar. For condensed matter, that would have been either Leonid Keldysh (of Keldysh Green functions), mostly on semiconductors, or the "Theory of Superconductivity" seminar, which *de facto* included all metal physics, and was moderated for many years by V.L. himself, and later by Kirzhnits.

V.L.'s rule was that no manuscript be submitted for publication before the work was presented at a Department's seminar. The seminars also included a "Journal Club" component: the first 15-20 minutes were dedicated to current literature and to travel reports. Nobody was supposed to have traveled to a conference without sharing afterward the highlights with the rest of the gang. The "Superconducting" seminar would usually start with V.L. distributing photocopies of selected current publications, mostly PRL and PRB. In the 1970's, photocopy machines were feared by the authorities more than firearms and probably justifiably so, for they could be used to disseminate banned literature. Ordinary people had no access to any copying equipment and so, although the Lebedev library subscribed to all Phys. Rev. journals, getting a copy of an article was complicated. V.L., as a full member of the Academy, had among numerous other privileges (starting with the right to shop at a special grocery), a right to request tables of contents of a number of journals from the central library of the Academy of Sciences with an option to order copies of any article. The copies would arrive in a few weeks, and (several months later than our Western colleagues) we would enjoy reading about the recent advances in the solid state physics. After browsing his daily mail, V.L. would select half a dozen of the most exciting articles and distribute them at the next seminar among the grad students, as well as staff scientists, requesting a brief account a week later. I vividly recall heated discussions that some of these articles induced.

I successfully defended my M. Sc. thesis in 1977, the same year the book [16] appeared in print. V.L.'s enthusiasm was still high, but the other members of the "High-$T_c$ task force" were gradually creeping away. Khomskii defected from superconductivity back to correlated magnetic oxides, Kopaev moved to other aspects of the excitonic insulator, unrelated to superconductivity, etc. Probably only V.L. remained firmly confident that HTSC would come some day; all others, to various degrees, were getting used to the idea that it was just a sweet dream.

Enter Aleksander "Sasha" Rusakov. Sasha was a good friend of mine, and a curious character on his own, probably deserving another article dedicated to his turbulent path from a juvenile detention facility to a university student and devoted Party member to a physicist and fervent anti-communist. At the stage in his life when we first met he had one love in life, and that love was called copper chloride. This thunderbolt, using the *Godfather* parlance, hit him around 1974 when he was an exchange visitor in the U.S., working with Paul Chu and Ted Geballe (I believe that he had started his first experiments with CuCl prior to that, in Moscow, but I am not sure). What he observed was a sudden drop of resistivity and some traces of diamagnetism under pressure at temperatures up to 90 K[23]. I know from the horse's mouth that Sasha already then believed he was seeing superconductivity, but apparently he had hard time selling this extravaganza to his older and more experienced advisors (Ref. 23 has no mention of superconductivity). Having come back from the States (which, as he confided to me at that time, he only did because his baby daughter remained a hostage in the Soviet Union), he vigorously pursued CuCl under pressure, this time mostly using a high pressure apparatus at the Moscow State University. The diamagnetic anomaly came back again and again, but Sasha and his collaborators at MSU could not achieve any sustainable reproducibility. His main senior contact at MSU, N.B. Brandt, got tired of waiting and insisted that even if the results were not routinely reproducible, they were exciting enough and should have been made public. He insisted on publishing them, this time with a clear claim of superconductivity at 90 K[24].

The JETP paper was an enormous hit. Sasha was immediately invited to give a two hour long presentation at a special session of the Academy of Science (which Ginzburg chaired, of course) and awarded extravagantly large funding for superconductivity research, specifically CuCl. His American collaborators, still skeptical, decided to publish a joint paper[25] on the results obtained during Sasha's visit to the States. Other groups joined the quest, and Sasha himself was able to procure the best possible equipment. Yet, the better the samples that Sasha and others worked with, the harder it was to observe the elusive diamagnetism! Initial enthusiasm gave way to general skepticism – with two exceptions: Sasha Rusakov and V.L. . At some point V.L. and Lev Gor'kov made a bet for a case of Cognac on whether "Rusakov's effect" was indeed superconductivity. To the best of my knowledge, the bet was never settled, although now, with the benefit of hindsight, it seems like Rusakov was seeing real high-Tc superconductivity in some impurity cuprate phase, which only formed in *bad* samples and only in a tiny amount.

As the CuCl hype was fading away, even V.L. seemed to be losing some of his faith. That was, of course, about the time the cuprates made their blatant appearance, cruelly smashing all stereotypes and prejudices, and indeed our entire idea of how science is made (inadvertently and indirectly leading to things like the Schön case). The bandwagon was overloaded and stayed so for at least a decade. My American contemporaries remember the legendary session at the APS March meeting that dragged on till morning. Though not quite at the same scale, we held a similar session of the Academy of Science of the USSR, with TV monitors in the street and so on.

I daresay, I had never before seen a man so completely happy and fulfilled as V.L. was in those days. Not all of us have an overwhelming life dream, and those who do more often than not fail to see it come true. V.L. has lived to see his, and if there is a Designer, he revealed his intelligence by granting this to V.L. I remember how in those days a conference speaker would show up at the morning session with red eyes, waiving dirty sheets of graph paper documenting another great discovery, "just observed last night". The tolerance threshold for dirty experiment and meaningless theories suddenly dropped to nearly zero, if not to a negative value.

Curiously, V.L., so radical in his aspirations before, was willing to accept much more conventional physics in HTSC than the newly formed fashion mandated[26]. Heated discussions in Ginzburg's Superconductivity seminar were always balanced, open to practitioners of the Hubbard model, as well as to those trying to address more conventional aspects of the problem. Band structure calculations were not considered a crime and the Fermi liquid was not being sent to the trash bin of history. On the contrary, V.L. encouraged me, as a relatively young (four years after Ph.D.) scientist involved with electronic structure, to write a short overview of the electronic structure of high-Tc cuprates for the Russian review magazine, Soviet Physics – Uspekhi. This overview, entitled *Electronic structure of high-temperature superconductors in normal state*, was published in 1989[27]. While there were no particularly deep insights there, I am proud to say (and not every theorist can say the same), that I do not need to take back any of the statements I made there, and I stand by my main conclusion, that the electronic structure is qualitatively different in undoped and optimally doped (using the modern jargon) cuprates, so that the main features of the electronic structure of the parent compounds are the Hubbard bands, while the electronic structure of fully doped materials will likely be reasonably well describable by the notion of the Fermi surface and band structure calculations (after an appropriate mass renormalization). I should remind the readers that initially the Fermi surface was missing in the photoemission measurements, with a number of theories building upon this fact. Later, as we all remember, experiments detected the Fermi surface, but not the bilayer splitting, again enkindling ingenious theories, only to be extinguished by newer experiments that conformed with band structure calculations.

Shameful as it is, now, twenty years later, while we congratulate V.L. with his 90 birthday, we cannot offer him as a birthday present the theory of high-temperature superconductivity. We do know that the simplistic, straightforward generalizations of BCS theory onto electron-mediated pairing, as discussed by Ginzburg, Little, Bardin and other great scientists in the sixties are most likely not relevant for the actual HTSCs as we have them now. Yet, the persistence, optimism, and physical intuition that kept Vitaly Lazarevich trusting that everything not forbidden by the laws of physics would eventually materialize should be a role model for the generation of physicists to come, as they were to our generation.

---

[1] I am using here and below the American terminology that roughly corresponds to the former Soviet system. Of course, there were no M. Sc. theses, but a so-called "diploma", no Ph.D., but a "candidate degree", no declaration of major in American sense, etc., but, in the first approximation, using the American terminology should give the reader a proper impression.

[2] Ironically, the same argument that was used (unofficially) to justify discrimination against Jews is now being used in the States to justify the Affirmative Action: Jews (meaning people whose Jewish ancestry was indicated in their state IDs, not the followers of the Jewish faith) comprised less than 2% of the population, but a much larger fraction of all college students, so for the sake of proportional representation they were discriminated against in the admission process.

[3] Usage of Russian patronymic names is not straightforward for an English speaker. In modern English, there are essentially two way of addressing a person: a more formal one, Mr./Dr./Ms. Doe, and a more intimate one, John/Mary etc. In Russian one has a choice of using (i) the second person singular or plural (like French *tu* and *vous*), (ii) the first name solely OR with the patronymic or (iii) using a salutation, Comrade or Citizen in the Soviet time, Gospodin (Master) now. An example can illustrate the basics of it: at Lebedev, my thesis adviser was Eugene "Jenya" Maksimov, about 15 years my senior and I also collaborated with Oleg Dolgov, a junior staff researcher at that time. The head of Superconductivity Section was Professor D.A. Kirzhnits, who will appear later on these pages. I was communicating with these people on a daily basis, addressing them, respectively, as Jenia + "vous", Oleg + "tu", and David Abramovich (of course, with a "vous").

[4] See, e.g., P.W. Anderson, Science, **144**, 373 (1964).

[5] M. L. Cohen and P. W. Anderson, "Comments on the maximum superconducting transition temperature," in *Superconductivity in d- and f-band Metals*, ed. D. H. Douglass (AIP, New York, 1972), p.17.

[6] D.A. Kirzhnits, E.G. Maksimov, and D.I. Khomskii, Lebedev Physical Inst. preprint #108 (Moscow, 1970); English translation: J. Low Temp. Phys. **10**, 79 (1973).

[7] N.N. Bogoliubov, V.V. Tolmachev, and D.V. Shirkov, *A New Method in the Theory of Superconductivity*, Consultants Bureau, 1959.

[8] P. Morel, P. W. Anderson, Phys. Rev. **125**, 1263 (1962). Incidentally, it appears to be one of very few papers in theoretical physics bearing the affiliation *French Embassy, New York, New York*.

[9] Anderson and Morel did not take into account the mass renormalization effects and derived $T_c(max)=E \exp(-3/\lambda)$, 50 times too high, but then proceeded to make an assumption of $\lambda<1/2$ (in fact, the largest known electron-phonon coupling constant is >2, in Pb-Bi alloys), obtaining $T_c(max)\sim 10$ K.

[10] O. V. Dolgov, D. A. Kirzhnits, and E. G. Maksimov, Lebedev Physical Inst. preprint #278 (Moscow, 1978); English translation: Rev. Mod. Phys. **53**, 81 (1981), see especially the Appendix: *The sign of the static dielectric function and the problem of superconductivity*.

[11] Although it is not the subject of this article, I cannot avoid mentioning in passing that D. Kirzhnits' name is practically unknown in the West as opposed to many other Soviet theorists of his generation, yet his understanding of physics and his command of the mathematical methods were second to none. The fact that he left a relatively modest trace in the memory of physics community is to a large extent due to his unorthodox, in fact, orthogonal to nearly all of us, idea of what constitutes an interesting and worthy problem in physics. For instance, he spent several years solving, brilliantly, the problem of the interaction of a magnetic monopole with matter, and I have yet to meet another person who would take interest in this problem.

[12] D. A. Kirzhnits, Usp. Fiz. Nauk **119**, 357 (1976); (Eng. tr. Sov. Phys.-Usp. **19**, 530).

[13] It is sometimes confusing to speak about the sign of pairing interaction without specifying, whether the sign change occurs in real or reciprocal space. In this quote, Anderson speaks about the real space, but an interaction can be repulsive everywhere in the reciprocal space and still be pairing (but not in the s-channel).

[14] M. E. Zhitomirsky, T. M. Rice and V. I. Anisimov, *Nature* **402**, 251 (1999)

[15] Yu. A. Uspenskii, Zh. Eksp. Teor. Fiz. **76**, 1620 (1979) (Eng. tran.: Sov. Phys. JETP, **49**, 822.

[16] V.L. Ginzburg, D.A. Kirzhnits (Eds) *Problema Vysokotemperaturnoi Sverkhprovodimosti (The Problem of High-Temperature Superconductivity)* (Moscow: Nauka, 1977) [Translated into English: *High-Temperature Superconductivity* (NewYork: Consultants Bureau, 1982)].

[17] E.A.Pashitski,ZhETF, 64, 2387 (1968); H. Frohlich, Proc. Phys. Soc. C1, 544 (1968).

[18] B. T. Geilikman, Sov. Phys. Uspekhi, 88, 327(1966).

[19] A piece of students' folklore goes that V.L. once failed in a qualifying exam a Ph. D. student working in quantum field theory because he did not know the BCS formula.

[20] Unfortunately, the flow of time makes us more and more specialized. The time of the Chaucerian *DOCTOUR OF PHISIK,* learned in every *magyk natureel,* from *astronomye* to *surgerye*, is long gone. I may be old-fashioned, but I am a little scared of experts in the 1D Hubbard model or pseudopotential calculations, with no interest in anything beyond their narrow field of expertise.

[21] Some remarks about the Soviet Physics seminars are here in place. As opposed to Western seminars, usually *ad hoc* gatherings on occasion of a visitor's talk, Russian tradition mandated regular, usually weekly seminars, more often two hours than one hour long, with an option of inviting a guest speaker, but often just one of the insiders reporting on a recent work. Seminars were usually known by the name of the moderator. V.L. himself orchestrated the work of three seminars, on general theory, on superconductivity and on astrophysics.

[22] The Lebedev Physical Institute had restricted access. The security regulations were similar to the place where I work now, the U.S. Naval Research Laboratory, although less severe than at most DOE labs (the Soviet analogue of which was Kurchatov Institute). Visits had to be reported to the security office by the hosting scientist, and the office would issue a day pass. Obviously the hundreds or more outside participants of the Ginzburg's Seminar (as it was universally known) could not physically be processed by one or two security officers, and V.L. had managed to procure a special arrangement that every Wednesday morning two of his grad students would sit at the gate, check IDs and write down the names of all seminar guests in one list that would be submitted to security *a posteriori*. For a while, I was one of those students.

[23] C. W. Chu, S. Early, T. H. Geballe, A. Rusakov and R. E. Schwall, J. Phys. C: Solid State Phys. **8**, L241 (1975)

[24] N. B. Brandt, S. V. Kuvshinnikov, A. P. Rusakov, and M. V. Semyonov, JETP Lett. **27**, 37 (1978); the authors' ordering is alphabetical.

[25] C. W. Chu, A. P. Rusakov, S. Huang, S. Early, T. H. Geballe, C. Y. Huang, Phys. Rev. **B 18**, 2116 (1978).

[26] A friend of mine told me a story about how he was having lunch during the first $M^2S$ conference in Interlaken, and two young fellows asked his permission to join his table. The parties introduced themselves, and the young physicists politely asked my friend, not much older than themselves at that time, what his field of research was. After hearing that he was doing band structure calculations, they blushed as if he admitted to pedophilia or secret membership in a neo-Nazi cult, and at the first possibility excused themselves and moved to another table.

[27] Soviet Physics - Uspekhi, **32**, 469-472 (1989).